\documentclass[twocolumn]{aastex62}
\usepackage{natbib}
\usepackage{color} 
\usepackage{aas_macros} 
\usepackage{hyperref} 
\usepackage{comment}
\usepackage{rotating} 

\newcommand{\kms}            {\,{\rm km\,s^{-1}}}

\received{2026}
\revised{}
\accepted{}

\begin{document}

\title[Galactic Stellar Halo Luminosity Function]{Galactic Stellar Halo Luminosity Function}

\author{Sarah A. Bird} 
\affil{College of Science, China Three Gorges University, Yichang 443002, People's Republic of China} \affil{Center for Astronomy and Space Sciences, China Three Gorges University, Yichang 443002, People's Republic of China}

\author{Chris Flynn} \affil{Centre for Astrophysics and Supercomputing, Swinburne
  University of Technology, Post Office Box 218, Hawthorn, VIC 3122,
  Australia}

\author{Rudra Sekhri}\affil{Centre for Astrophysics and Supercomputing, Swinburne
  University of Technology, Post Office Box 218, Hawthorn, VIC 3122,
  Australia}
  
\author{Hai-Jun Tian} \affil{School of Science, Hangzhou Dianzi University, Hangzhou 310018, People's Republic of China}
\affil{Big Data Institute, Hangzhou Dianzi University, Hangzhou 310018, People's Republic of China}

\author{Juntai Shen} 
\affil{Department of Astronomy, School of Physics and Astronomy, Shanghai Jiao Tong University, 800 Dongchuan Road, Shanghai 200240, People's Republic of China}
\affil{Key Laboratory for Particle Astrophysics and Cosmology (MOE) / Shanghai Key Laboratory for Particle Physics and Cosmology, Shanghai 200240, People's Republic of China}

\author{Xiang-Xiang Xue}
\affil{CAS Key Laboratory of Optical Astronomy, National Astronomical Observatories, Chinese Academy of Sciences, Beijing 100101, People's Republic of China}
\affil{Institute for Frontiers in Astronomy and Astrophysics, Beijing Normal University, Beijing, 102206, People's Republic of China}

\author{Chao Liu}
\affil{Key Laboratory of Space Astronomy and Technology, National Astronomical Observatories, Chinese Academy of Sciences, Beijing 100101, People's Republic of China}
\affil{Institute for Frontiers in Astronomy and Astrophysics, Beijing Normal University, Beijing, 102206, People's Republic of China}
\affil{School of Astronomy and Space Science, University of Chinese Academy of Sciences, Beijing 100049, People's Republic of China}

\author{Gang Zhao}
\affil{CAS Key Laboratory of Optical Astronomy, National Astronomical Observatories, Chinese Academy of Sciences, Beijing 100101, People's Republic of China}
\affil{School of Astronomy and Space Science, University of Chinese Academy of Sciences, Beijing 100049, People's Republic of China}

\email{sarahbird@ctgu.edu.cn}

\begin{abstract}

We measure the luminosity function (LF) of the Milky Way's stellar halo, using a magnitude complete, distance limited sample of stars from {\it Gaia} DR3. Stars with high transverse velocities are selected, to isolate a high purity sample of the local halo. We adopt a cutoff transverse velocity of 250 $\kms$, yielding 24,471 stars, and compute the halo LF, taking into account the effects of sample selection criteria. The LF displays similar features as are found in the well-probed LF of nearby, metal-rich disk stars, showing a strong peak at an absolute magnitude of around $M_G=10$, and a flattening near $M_G\sim7$ (Wielen dip). The {\it Gaia} sample yields the first measurement of the LF continuously from the dimmest main sequence halo stars (subdwarfs) at an absolute $M_G$ magnitude near 13 mag to bright giants at $M_G\sim-3$. 
We obtain a local stellar halo number density of $1.7\times10^{-4}$ stars\,pc$^{-3}$ 
and disk-to-halo ratio by stellar number density
of 480:1. 
We convert the {\it Gaia} $G$ band measurements for our sample stars to Johnson-Kron-Cousins $V$ band, compute the $V$-band halo LF, and compare it to previous studies published over many decades that cover a wide range of techniques used. We discuss applications of the LF to the measurement of the luminosity and stellar mass of the Milky Way halo. 

\end{abstract}

\keywords{galaxies: individual (Milky Way) --- 
Galaxy: halo ---
Galaxy: kinematics and dynamics ---
Galaxy: stellar content --- 
stars: kinematics and dynamics}

\section{Introduction} \label{sec:intro}

Stellar luminosity functions (hereafter LFs) are a fundamental property of stellar populations in Galactic astronomy. The function is a measurement of the number of stars per unit volume as a function of absolute magnitude;
in other words, the number density of stars per luminosity interval. It has a wide range of uses, from determining the evolutionary stages and ages of a population of stars, calibrating star count models for the structure of the Milky Way, testing theoretical models of stellar structure and evolution, conversion to the mass function of stars, understanding integrated light of external galaxies, among many others.

The measurement of the luminosity function of disk stars in the solar neighborhood has a long history \citep[{\it e.g.},][]{Kapteyn1902,Wielen1983,Jahreiss1997,Reid2002,Just2015,GaiaCollaborationSmart2021}. Its measurement bridges the eras of ground-based parallax measurements in programs at many observatories through most of the 20th century, to the era of space-based astrometry with the European Space Agency (ESA) {\it Hipparcos} (1989$-$1995) and ESA {\it Gaia} missions (2020$-$2025).  During this period, the seminal ground-based parallax work
was compiled into the Gliese Catalogue \citep{Gliese1957}, later termed the Catalogue of Nearby Stars \citep[CNS,][]{Gliese1957,Gliese1969,Gliese1991} and later still combined with space-based astrometry \citep{Jahreiss1997,Golovin2023}. Complete parallax catalogs of magnitude-limited, space-based astrometry followed, the ESA {\it Hipparcos} Catalog \citep[][]{ESA1997} and most recently the {\it Gaia} Catalogue of Nearby Stars \citep{GaiaCollaborationSmart2021}. Disk stars represent more than 99 per cent of stars within short (5$-$200 pc) distances of the Sun \citep[{\it e.g.},][]{Dawson1986,Cooke2000,Reid2005}, and the most recent measurement of the local disk LF by {\it Gaia} is highly accurate over a range of absolute magnitudes from $-1.0$ (representing high mass upper main sequence stars and evolved stars on the giant branch) to $20.5$, approaching the hydrogen burning limit \citep{GaiaCollaborationSmart2021}.   

The luminosity function of the stellar halo on the other hand has yet to be well measured consistently over the very wide range of luminosity as has been attained for disk stars. Halo stars are well known to differ markedly from disk stars in being much older, significantly more metal poor, and having very different kinematics within the Milky Way. They are rare, comprising less than 1 per cent of the stars near the Sun. Estimates for the local ratio of disk-to-halo stars range from $\sim200\simeq900$ locally to the Sun \citep[{\it e.g.},][]{Dawson1986,Morrison1993,Cooke2000,Reid2005}, with the value of this ratio varying from study to study owing to the uncertainties in modeling and differentiating the disk and halo populations. 

Because of the rarity of halo stars, efficient techniques for finding them have been developed over many decades of sky surveys. In the early twentieth century, the use of proper motions (and the closely related ``reduced proper motions,'' RPM) proved an effective way of finding halo stars, leveraging their very different kinematics and somewhat different luminosities compared to disk stars. By mid-century, it was found that halo stars are metal-poor, significantly affecting their colors and luminosities on both the main sequence and later evolutionary stages, so that `chemistry/abundances' could be used to find them. This lead to halo stars being isolated via `color excess' methods (particularly in the $UV$/blue) and/or by spectroscopic methods in targeted surveys. The intrinsic differences of these methods compared to the proper motion-selected samples allowed the measurement of kinematically unbiased halo samples from which the kinematics of the halo became better understood. An era of very significant surveys for halo stars in the late twentieth century yielded important insights into the early evolution and assembly of the Milky Way, through the kinematics, metallicity distribution and abundance patterns of tens of thousands of such stars \citep[see reviews by, {\it e.g.},][]{Freeman1987,Beers2005,Helmi2008,Ivezic2012}.

Using these varying methods to collect halo star samples, many authors have measured the stellar halo LF over the decades; these works are reviewed by \citet{Mould1982, Bahcall1986araa, Bessel1993, Mould1996, Reid2000, Reid2005}.

One would most like to have a volume-complete samples from which to measure LFs. With these things in mind, we are now in prime position with {\it Gaia} to attain a volume-complete sample, homogeneously covering a large range of magnitudes, providing distances and velocities from which we can select halo stars. The simultaneous provision of parallaxes, proper motions, and for a significant fraction of the samples, radial velocities, means that 5D or 6D astrometric information, surveyed uniformly over the whole sky, can allow kinematic methods to cleanly isolate very pure samples of the stellar halo. 

Updates to the Galactic stellar halo LF have remained absent for nearly two decades! But we now can measure it in light of the latest data coming from {\it Gaia} DR3. In Sect. \ref{sec:data} we present the data, the selection of halo stars using tangential velocities, and the method and corrections used to measure the LF. In Sect. \ref{sec:lf}, we present our resulting stellar halo LF.
We discuss and compare our newly measured LF to the decades past of previous determinations of the LF, ours for the first time using the largest sample that covers the widest range from bright giants to faint main sequence stars (Sect. \ref{sec:discussion}). We draw our conclusions in Sect. \ref{sec:conclusion}.

We note that we exclude white dwarfs in our current study, and refer the reader to the work of \citet{HarrisHugh2006}, \citet{Rowell2011}, and \citet{QiuTian2024} who measured the LF for nearby halo white dwarfs.

\section{Stellar Halo Sample from {\it Gaia} DR3 Data} \label{sec:data}

We select our halo star sample from {\it Gaia} DR3 \citep{GaiaCollaborationPrusti2016, GaiaCollaborationVallenari2023, Babusiaux2023}. The initial data set is selected primarily on a distance upper limit, an apparent magnitude cut, and by Galactic latitude. We select halo stars using a tangential velocity criteria. We describe these criteria fully in Sect.\,\ref{sec:dataselect} and \ref{sec:haloselect}. In Sect.\,\ref{sec:complete}, we describe the completeness corrections we apply to our results due to our halo sample selection criteria.

\subsection{{\it Gaia} DR3 Data Selection} \label{sec:dataselect}

Our initial data sample of sources within a sphere of 1\,kpc radius about the Sun yields $39\,669\,995$ sources. See Appendix\,\ref{app:adql} for details on this selection.

We further refine the initial sample with the following criteria.

To avoid the extinct regions of the thin dust layer in the disk, we select stars with Galactic latitude $|b|>36^\circ$ following \citet[][see their figure 1]{Dixon2023}.

We adopt a limiting magnitude of $G=17$\,mag, motivated by the completeness analysis of the DR2 release \citep{GaiaCollaborationBrown2018}. We tested this with mock catalogs and found it to be an appropriate $G$ band limit for DR3 as well.

In Appendix \ref{app:GV} we describe converting {\it Gaia} $G$ to Johnson-Kron-Cousins's $V$; when using our calculated $V$ magnitudes, we adopt a limiting magnitude of $V=18$\,mag. 

In the following Sect.\,\ref{sec:haloselect}, we describe our tangential velocity criteria to select halo stars as well as the removal of halo white dwarfs. We give our final sample size at the end of Sect.\,\ref{sec:haloselect}.

\subsection{Halo Star Selection Criterion: $V_\mathrm{t}>250\kms$} \label{sec:haloselect}

Our primary selection criterion for the stellar halo is the transverse velocity. With the wealth of {\it Gaia} data, we are at a privileged position to have high quality proper motions ($\mu_\alpha,\mu_\delta$) (mas\,yr$^{-1}$) and parallaxes $\varpi$ (mas) for an unprecedentedly large sample. The parallaxes and proper motions are used to calculate the transverse velocity $V_\mathrm{t}$

\begin{equation}
V_\mathrm{t}=4.74\sqrt{\mu_\alpha^2+\mu_\delta^2}/ \varpi. 
\end{equation}

As is well known, halo stars have high velocity dispersions and also a high asymmetric drift relative to disk stars. Consequently, high transverse velocities $V_\mathrm{t}$ have high discriminatory power in isolating halo stars from the much larger numbers of disk stars in the sample volume and has been used over several decades \citep[{\it i.e.},][]{Bahcall1986,Dahn1995}. This method requires that all, or at least a substantial fraction of the stars have parallaxes. This method is much more efficient at isolating the halo than using proper motions (or reduced proper motions) alone. 

The effect of transverse velocity as a population discriminator within the color-magnitude diagram (CMD) has been shown for Gaia DR2, \citet{GaiaCollaborationBabusiaux2018} (see in particular figures 21 and 22). The CMDs of stars predominately of the disk with $V_\mathrm{t} < 40 \kms$, the thick disk ($ 60 < V_\mathrm{t} < 150$) and the halo ($V_\mathrm{t} > 200$) show substantially different CMDs, reflective of their different age and abundance distributions.  

An important correction needs to be applied for the tangential velocity method, which accounts for halo stars which overlap in $V_\mathrm{t}$ with the old disk and thick disk. Simulations of the kinematics of these populations, {\it e.g.}, \citet[][]{Bahcall1986} have shown that cuts $V_\mathrm{t} > 200$ to 250 $\kms$ yield high purity halo stars samples, at the expense of excluding up to approximately half the stars. This incompleteness can be corrected for quite effectively, by modeling the missing fraction based on the kinematic parameters of the Milky Way's stellar populations, and the survey selection cuts \citep{Bahcall1986}. 

\begin{figure*}
\includegraphics[]{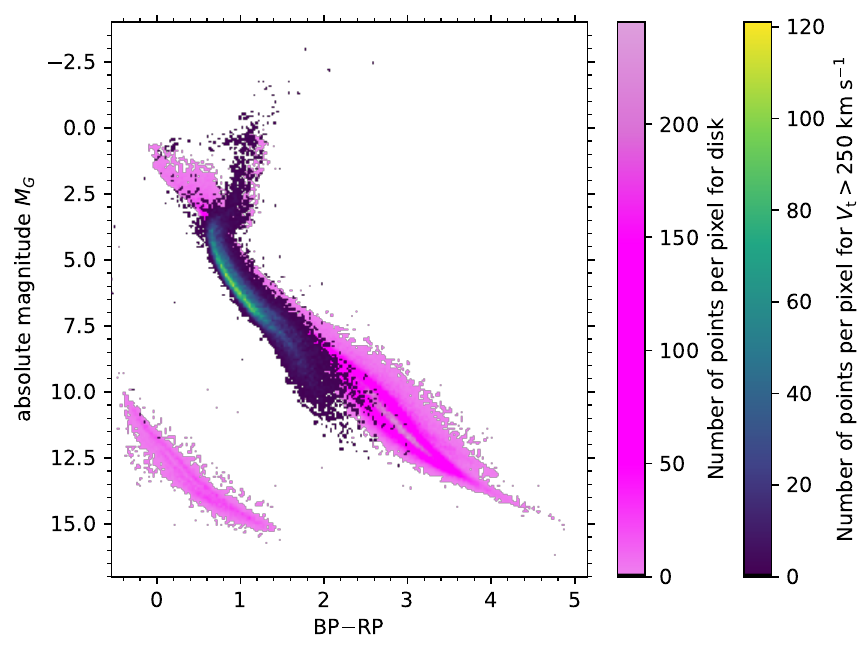}
\caption{Color-magnitude (Hertzsprung-Russell) diagram in {\it Gaia} passbands absolute magnitude $M_G$ and color BP$-$RP.
The viridis-like-colored scatter-density plot represents the halo stars with $V_\mathrm{t}>250$\,km\,s$^{-1}$, $1/\varpi<1$\,kpc, $|b|>36^\circ$, and $G<17$. The underlying magenta scatter-density plot maps a representative disk sample with $V_\mathrm{t}<40$\,km\,s$^{-1}$ and $1/\varpi<0.1$\,kpc. Colorbars indicate the number counts.
} 
\label{fig:cmd}
\end{figure*}

By making a simple and classical proper motion cut \citep[{\it e.g.},][]{SchmidtMaarten1975,Bahcall1986} the retention of halo stars from the proper motion-limited sample leads to a disk-to-halo ratio of sample stars as low as $\sim4:1$ \citep[{\it e.g.},][]{Cooke2000,Digby2003}. This classical type of proper motion selection criteria has most recently been used to define stellar halo samples from {\it Gaia} by, {\it e.g.}, \citet[][]{GaiaCollaborationBabusiaux2018,Gallart2019,Sahlholdt2019,Amarante2020a}. 
We thus select halo stars using high tangential velocities as the criterion.

The cutoff criterion between disk and halo is traditionally set quite high, in order to achieve high purity of the halo stars, but this can be at considerable cost of the sample size. Cutoff values in the literature range from 200$\,\kms$ \citep[{\it e.g.},][]{GaiaCollaborationBabusiaux2018} up to 250$\,\kms$ \citep[{\it e.g.},][]{SchmidtMaarten1975}. This brackets the value of the local standard of rest of $\approx 220\,\kms$, and is thus primarily selecting stars with retrograde rather than prograde orbits around the Galactic center, so that to first order, about half of the halo stars are missed. An important task is to estimate what this fraction is, so that the density of the stars can be appropriately corrected. The most important contaminant into the halo is the thick disk, which has sufficient velocity dispersion that the tail of the velocity distribution overlaps with the halo proper. In the analysis of a similar data set as we use, \citet{Amarante2020a} concluded that in a sample of stars with $V_\mathrm{t}>200\kms$ there remains 13 per cent of stars with thick-disk kinematics. They find that cutoffs of 220 and $250\kms$ retain $\approx 94$ and 98 per cent stars with stellar halo kinematics.

In Fig \ref{fig:cmd}, we divide the 
sample at $V_\mathrm{t}>250\kms$ to show the CMD of halo stars, overlaid on a representative sample of disk stars (magenta). We can clearly see the classical halo subdwarf, giant, and red and blue horizontal branches, and a scatter of likely blue stragglers and a few white dwarfs. The plot is dominated by subdwarfs (main sequence stars belonging to the halo) seen located below (less luminous) and to the left (bluer) of the disk main sequence. This is most clearly seen around $M_G = 6$, BP$-$RP$ = 1$. Halo giants lie somewhat to the blue of the disk giants (seen around $M_G = 1$, BP$-$RP$ = 1$).
Helium burning stars (`horizontal branch' stars) are seen at a range of colors $0 < $BP$-$RP$ < 1.5$ at $M_G \approx -1$ and there is evidence for `blue straggler' stars at $M_G \approx 3$, BP$-$RP$\approx 0.5$.  

We have used those stars selected in Sect. \ref{sec:dataselect} that have radial velocities measured by {\it Gaia} to create a kinematical model of the local Milky Way, with which we are able to quantify the effectiveness of the tangential velocity cut-off in isolating halo stars from the other components. We model the Milky Way using five components: young disk, thin disk, thick disk, underlying halo, {\it Gaia}-Enceladus-Sausage satellite. By comparing $V_\mathrm{t}$ of the model and of {\it Gaia} we determine the appropriate velocity cut that allows us to exclude the majority of the disk and thick disk contamination. A full description of this modeling is given in Appendix \ref{app:mwmodel}. We find that with $V_\mathrm{t}>250\kms$ we achieve a pure halo sample.

In Sect.\ref{sec:volumecomplete} we explain our corrections to the volume completeness of the LF and in Sect. \ref{sec:vtancomplete} we quantify the completeness correction needed due to our $V_\mathrm{t}$ cutoff.

We remove the halo white dwarfs from further study in this work using the criterion $M_G<7+4\times(\mathrm{BP}-\mathrm{RP})$.

Taking into account all these selection criteria, we attain 24,471 halo stars from which we measure the LF in {\it Gaia} $G$ and 29,428 in Johnson-Kron-Cousins $V$.

\subsection{Halo Star Completeness Corrections} \label{sec:complete}

\subsubsection{Volume Completeness Correction} \label{sec:volumecomplete}

The volume completeness of our sample is influenced by two primary factors, {\it i.e.}, our selection criterion in Galactic latitude of $|b|>36^\circ$ and the maximum distance that $Gaia$ can detect the stars of a given luminosity in the sample. Since we use a spherical volume we can directly integrate the appropriate volume of our sample due to the Galactic latitude cut and maximum distance. These corrections are relatively straightforward. The corrections for the tangential velocity cut are quite complex and are discussed next. 

\subsubsection{Tangential Velocity Completeness Correction} \label{sec:vtancomplete}

By selecting stars with transverse velocities $V_\mathrm{t}>250\,\kms$ we achieve a pure halo sample, but we also exclude half of the total number of halo stars with such a selection criterion as these halo stars have similar transverse velocities as disk stars. We need to correct our luminosity function for the number of halo stars lost due to our $V_\mathrm{t}$ cut.

To estimate the fraction of stars lost due to our $V_\mathrm{t}$ cut of 250\,$\kms$ we use the halo components of our Milky Way model (described in Appendix \ref{app:mwmodel}) to investigate the halo distribution function in 3D velocity space within the solar neighborhood using the two components, one representing the underlying halo and the second the {\it Gaia}-Enceladus-Sausage satellite \citep[][]{Belokurov2018,Deason2018.862,Haywood2018,Helmi2018,Koppelman2018}.

As detailed in Appendix \ref{app:mwmodel}, we find a correction factor of 2.0 and we apply this in our luminosity function calculations. Such correction factors to detect halo stars depend on the kinematic model and $V_\mathrm{t}$ cut applied \citep[{\it e.g.},][]{Richstone1981}. Several examples from previous works include $3.0\pm0.4$ \citep{Bahcall1986}, 2.01$-$3.70 \citep{Gizis1999}, and 1.72$-$2.04 \citep{Digby2003}.

\section{Halo Luminosity Function} \label{sec:lf}

\begin{figure*}
\includegraphics[width=2.1\columnwidth]{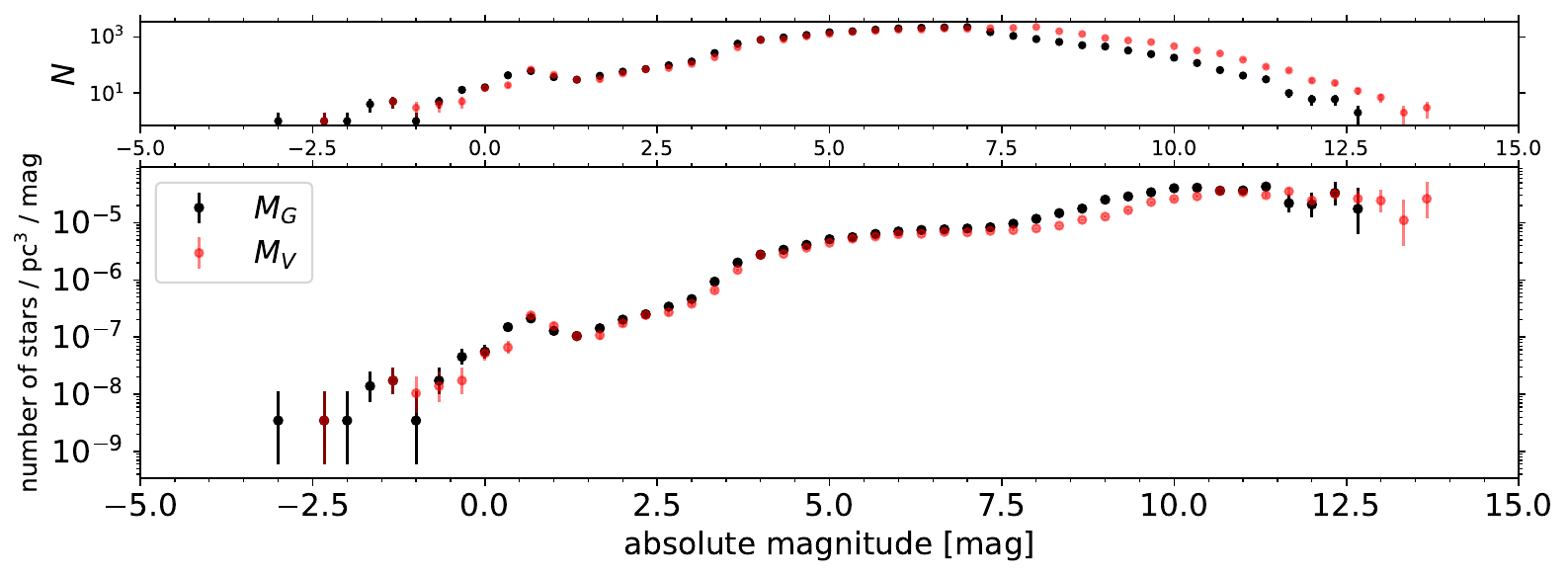}
\includegraphics[width=2.1\columnwidth]{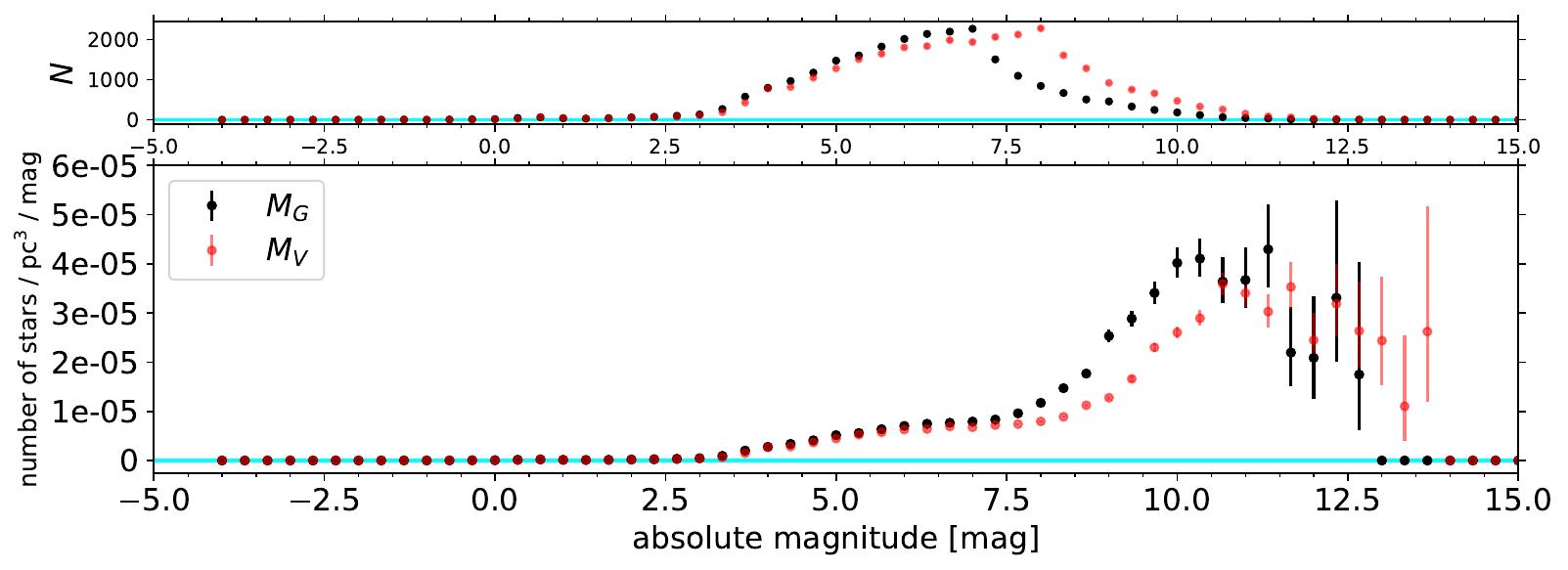}
\caption{
{\bf Uppermost panel:} Number of halo stars in bins of absolute magnitude $M_G$ (black markers) and $M_V$ (red markers) in our final sample.
{\bf Panel second from the top:} LF, that is, the local number density of stars per absolute magnitude ($M_G$ are black markers and $M_V$ red markers) for the stellar halo (excluding white dwarfs) in the solar neighborhood at $Z=0$ pc.
{\bf All panels:} The bin widths are 0.333 magnitudes in absolute {\it Gaia} $M_G$ magnitude and absolute Johnson-Kron-Cousins $M_V$ magnitude. Confidence intervals are Poisson uncertainties \citep{Gehrels1986}.
{\bf Second from the bottom and lowermost panels:} Same as the upper two panels; but the vertical axes are scaled linearly instead of logarithmically, and the horizontal cyan line denotes zero.
} 
\label{fig:lf}
\end{figure*}

\begin{figure*}
\includegraphics[width=2.1\columnwidth]{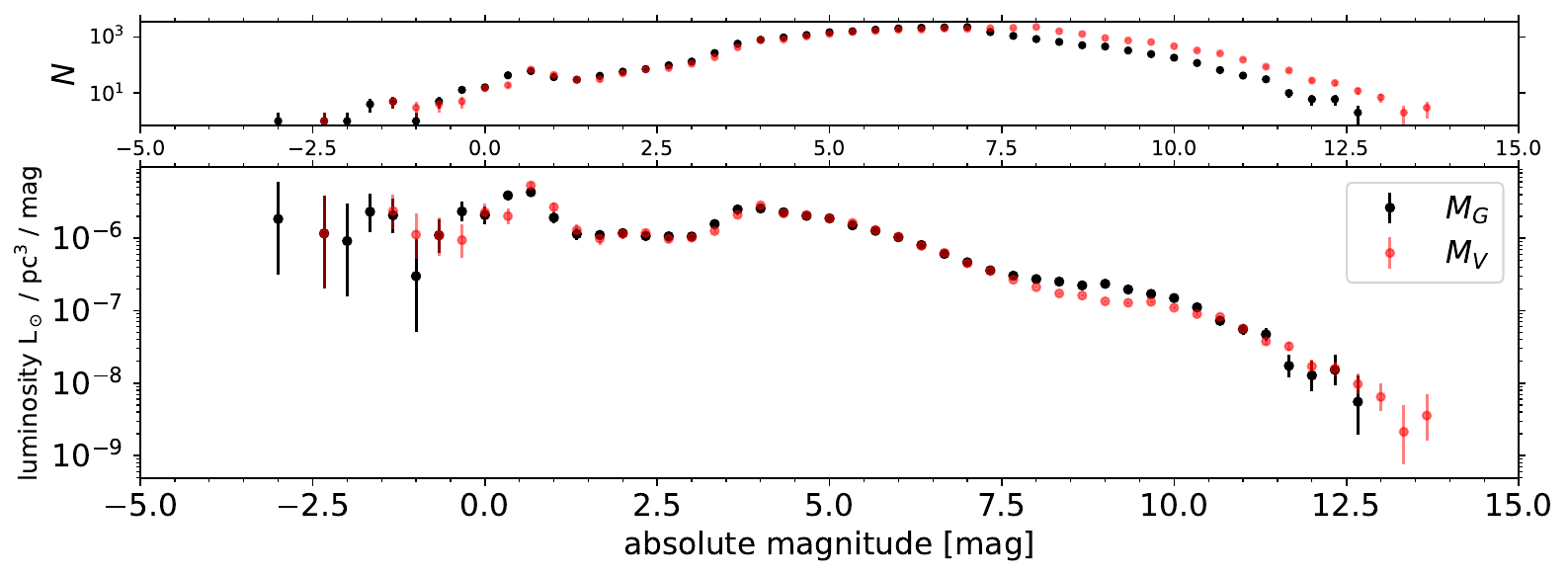}
\includegraphics[width=2.1\columnwidth]{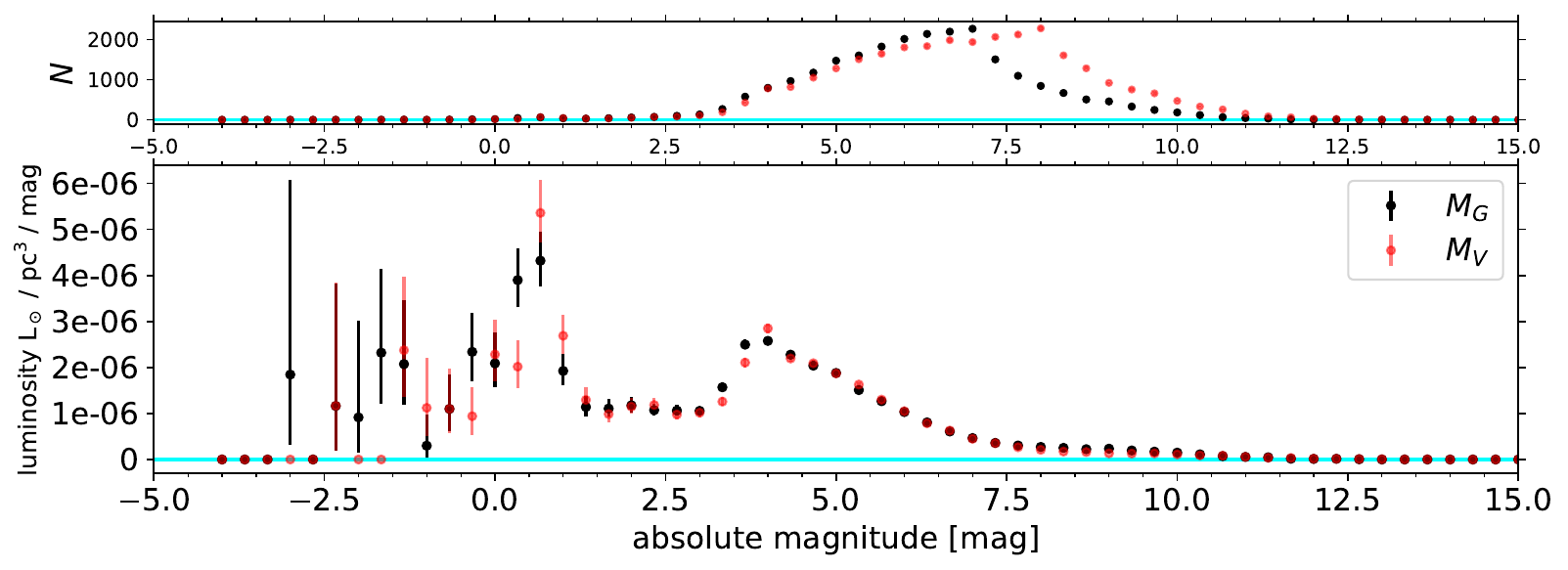}
\caption{
{\bf Uppermost panel:} Number of halo stars in bins of absolute magnitude $M_G$ (black markers) and $M_V$ (red markers) in our final sample.
{\bf Panel second from the top:} Luminosity density function, that is, the local luminosity density of stars measured in solar luminosities L$_\odot$ per cubic parsec per magnitude $M_G$ (black markers) and $M_V$ (red markers) for the stellar halo (excluding white dwarfs) in the solar neighborhood at $Z=0$ pc.
{\bf Both panels:} The bin widths are 0.333 magnitudes in absolute {\it Gaia} $M_G$ magnitude and absolute Johnson-Kron-Cousins $M_V$ magnitude. Confidence intervals are Poisson uncertainties \citep{Gehrels1986}.
{\bf Second from the bottom and lowermost panels:} Same as the upper two panels; but the vertical axes are scaled linearly instead of logarithmically, and the horizontal cyan line denotes zero.
} 
\label{fig:lumdensity}
\end{figure*}

\begin{figure*}
\includegraphics[width=2.1\columnwidth]{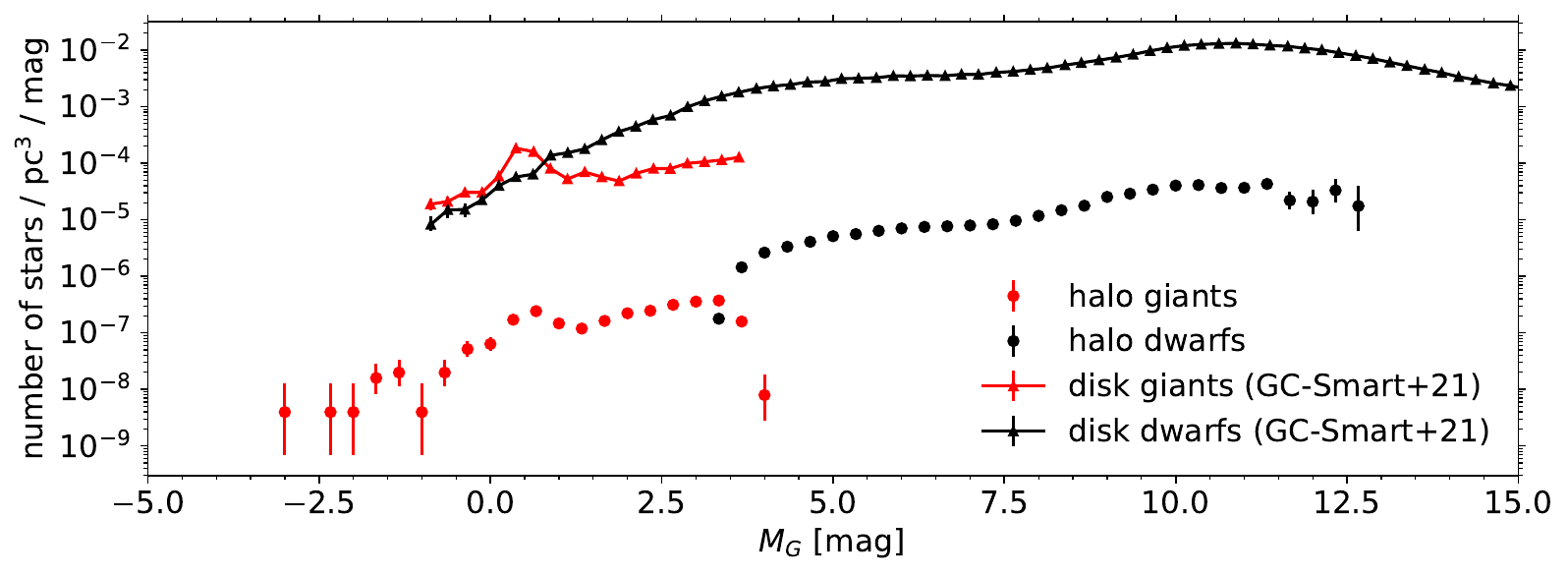}
\caption{
Luminosity function for the stellar halo (dot markers) within the solar neighborhood compared with the disk luminosity function \citep[triangles,][GC-Smart$+$21]{GaiaCollaborationSmart2021}. Main sequence dwarfs are marked in black, giants in red. 
The GC-Smart$+$21 disk LF extends to $M_G>20$, but here we plot only till the dimmest extent of the halo subdwarfs ($M_G<15$).
The bin widths are 0.333 (halo) and 0.25 (disk) magnitudes in absolute {\it Gaia} $M_G$ magnitude.
} 
\label{fig:lf-disk}
\end{figure*}

The second panel from the top of Fig. \ref{fig:lf} presents our resulting stellar halo LFs (scale of the vertical axis is logarithmic) using the absolute {\it Gaia} magnitude $M_G$ (black markers) and absolute Johson-Kron-Cousins magnitude $M_V$ (red markers) for the stellar halo sample excluding white dwarfs in bins of 0.333 magnitude. This measurement is the number density of halo stars per magnitude per cubic parsec in the solar neighborhood at $Z=0$ pc. The LFs have been corrected due to the selection criteria and completeness. The uppermost panel presents the number histogram for the halo stars in our sample used to measure the LF, a spherical volume with radius of 1 kpc. The majority of our bins have 10s to $>1000$ stars. We plot Poisson uncertainties \citep{Gehrels1986}; the large majority of our bins have error bars smaller than the marker. The upper two panels and the lower two panels differ only in the scale of the vertical axis, logarithmic and linear, respectively.

When plotting the LF using a logarithmically scaled vertical axis, a small peak appears centered near $M_G,M_V\sim0.7$; these are the horizontal branch stars. 
In a future effort we plan to increase our distance limit to, {\it i.e.}, 4 kpc as this will increase the number of intrinsically bright giants observable by {\it Gaia} such as sampled in the tip-of-the-red-giant study by \citet{Dixon2023}.

We also find in the halo LF other classical features found in the disk LF, {\it i.e.}, the Wielen dip \citep{Wielen1974,Wielen1983} with $M_G\sim5-9$ and the maximum of the LF that is reached by $M_G\sim10$. We find our stellar halo LF remains relatively flat towards dimmer magnitudes. The uncertainties begin increasing at dimmer magnitudes such that the form of the LF, whether remaining flat or clearly turning over, for $M_G>11.5$ remains uncertain. 
We further our discussion of features in the LF in Sect. \ref{sec:lffeatures}.

By integrating our stellar halo LF, we obtain a robust measurement of the local stellar halo number density, $1.7\times10^{-4}$ stars\,pc$^{-3}$
and $1.6\times10^{-4}$ stars\,pc$^{-3}$ measured from the $M_G$ and $M_V$ LFs, respectively. The relative difference between these local measurements is a small 6 per cent. We use the stellar density of the disk 0.081 stars\,pc$^{-3}$ from \citet{GaiaCollaborationSmart2021} to find a
disk-to-halo ratio by stellar density 
of 480:1 and 500:1 derived from our LFs measured in $M_G$ and $M_V$, respectively.

We make use of the estimated absolute magnitude of the Sun in $M_G$ and $M_V$, equal to  4.667\,mag (Vyas, et al., in preparation) and 4.81\,mag \citep{Willmer2018}, respectively, to calculate the local luminosity density function for both $M_G$ and $M_V$ measured in solar luminosity L$_\odot$ per cubic parsec per magnitude (Fig. \ref{fig:lumdensity}).
By comparing Figs \ref{fig:lf} and \ref{fig:lumdensity}, we can see that although the subdwarfs dominate the stellar number density, the giants are the main contributor to the luminosity density.

By integrating our local stellar halo luminosity density functions from Fig. \ref{fig:lumdensity} measured using $M_G$ and $M_V$, we obtain robust estimates for the total local stellar halo luminosity density at $Z=0$\,pc, finding $1.7\times10^{-5}$ L$_\odot$\,pc$^{-3}$ and $1.5\times10^{-5}$ L$_\odot$\,pc$^{-3}$ in the $G$ and $V$-band, respectively.

By assuming a density profile for the halo stars in our Galaxy, we can use our local number density and local luminosity density values as robust normalizations to estimate the total number of halo stars and the total luminosity of the stellar halo for the Milky Way.

We assume a broken power law for the stellar halo with $\alpha_\mathrm{inner}=-2.5$, $\alpha_\mathrm{outer}=-4.0$, and break radius at a Galactocentric radius of 20\,kpc. This Galactic stellar halo density profile is motivated by the results of many previous works \citep{Watkins2009,Sesar2010,Sesar2011,Sesar2013,Deason2011.416,Deason2014,Faccioli2014,Pila-Diez2015,Xue2015}. We estimate the luminosity in spherical shells of width 0.01\,kpc from a Galactocentric radius of 0.1 to 100\,kpc.

Using $M_G$ and estimating out to 100\,kpc, we find $4.6\times10^9$ Milky Way halo stars and a total Galactic stellar halo luminosity of $4.6\times10^8$\,L$_\odot$ and absolute magnitude of -17.0 mag in the $G$-band. Using $M_V$ and estimating out to 100\,kpc, we find $4.3\times10^9$ Milky Way halo stars and a total Galactic stellar halo luminosity of $4.1\times10^8$\,L$_\odot$ and absolute magnitude of -16.7 mag in the $V$-band.

In Fig. \ref{fig:lf-disk} we separate the halo main sequence dwarf (black dots) and giant (red dots) LFs and compare these to those of the disk (black and red triangles). The disk LFs are taken from \citet{GaiaCollaborationSmart2021} using {\it Gaia} eDR3.

We define stellar halo main sequence dwarfs dimmer than the giant branch as all stars (excluding white dwarfs) using $M_G>5.5$. For brighter main sequence dwarfs with $M_G<5.5$, we keep stars with $M_G>-8 \times ($BP$-$RP$)+8.3$.

For our stellar halo sample with $M_G>2$, we define giants by $M_G<4$ and $M_G<3.9 \times ($BP$-$RP$)+0.1$. We combine this with all halo stars having $M_G<2$. 

The {\it Gaia} eDR3 disk LFs from \citet{GaiaCollaborationSmart2021} are measured from stars located within 100 pc around the Sun. As we are collecting the halo stars from within 1 kpc around the Sun, we are able to sample the intrinsically bright stars up to $M_G\sim-3$ located at larger distances that are missed by the 100 pc sample as {\it Gaia} does not include stars brighter than apparent magnitude $G\simeq3$.

We plot the disk LF Poisson uncertainties given by \citet{GaiaCollaborationSmart2021}. As we pointed out concerning the halo LF, the uncertainties on the disk LF are smaller than the marker size for many bins.

We can see that there is only a small overlap at around $M_G = 3.5$ between the stellar halo main sequence stars in the LF, and those on the giant branch. This is because higher mass halo stars than those at the turnoff mass are evolving off the main sequence. On the other hand, the disk main sequence and the giant branch LFs overlap substantially. This is because the disk contains a mixture of younger, more massive stars still on the main sequence as well as giants with a range of ages and masses that have recently evolved off the main sequence. We make further comparisons of features in the LF in Sect. \ref{sec:lffeatures}.

After converting the stars to $V$-band using $G$ and BP$-$RP  (see Appendix \ref{app:GV}) we present in Fig. \ref{fig:lf-lit} the $V$-band halo LF along with many halo star LFs from the literature.
Overall we find good agreement although our stellar halo sample contains the largest amount of data and covers the largest span of magnitudes consistently observed from the same satellite. We make more detailed comparison between studies in Sect. \ref{sec:litcompare}.

\begin{figure*}
\includegraphics[width=2\columnwidth]{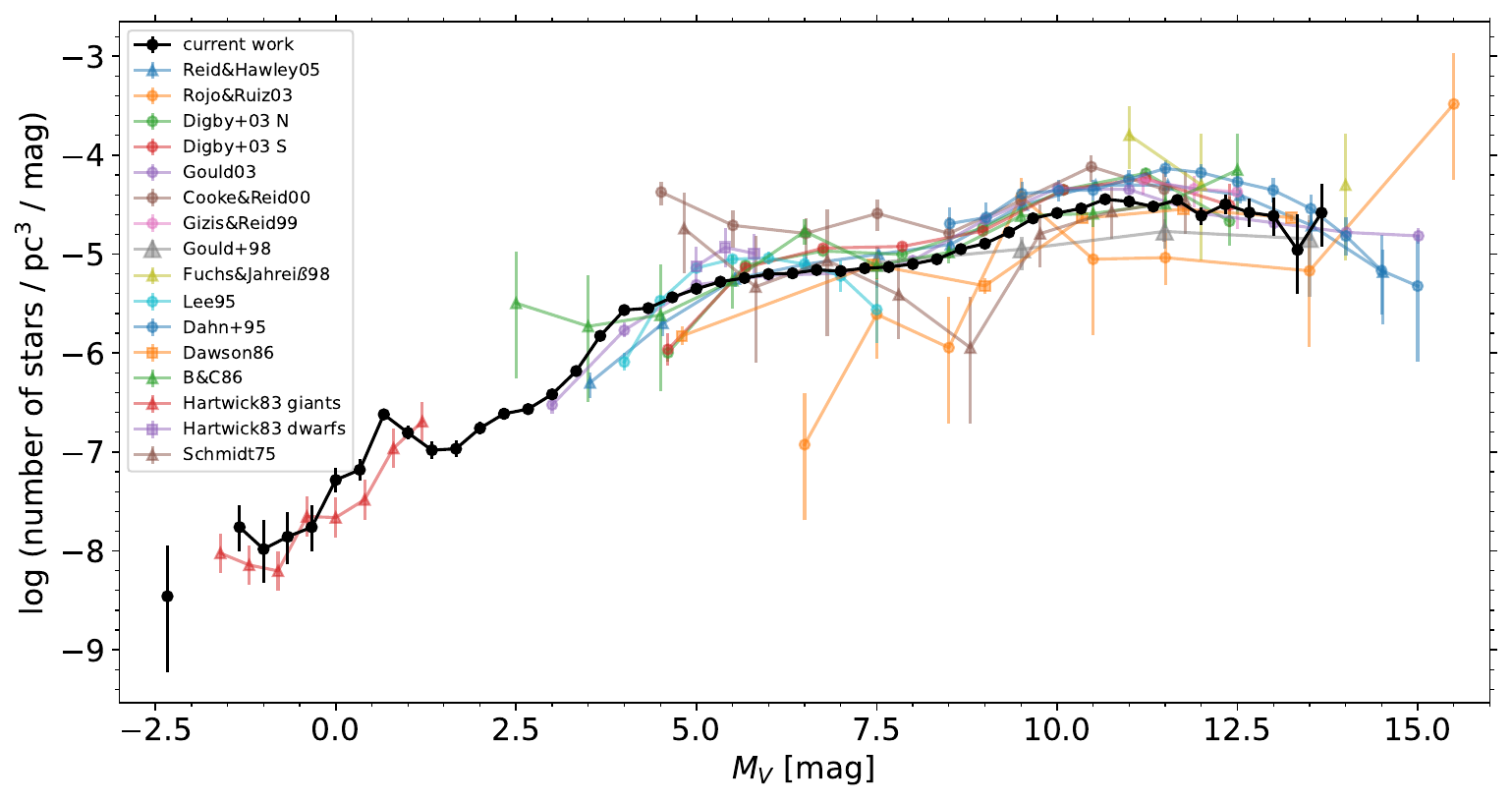}
\caption{Johnson-Kron-Cousins $M_V$ LF for {\it Gaia} halo stars within the solar neighborhood compared with stellar halo LFs from the literature. {\it Gaia} halo LF uses bin widths of 0.333 magnitude in $M_V$ and the confidence intervals are Poisson uncertainties \citep{Gehrels1986}. See text and Table \ref{table:lit} for details.
} 
\label{fig:lf-lit}
\end{figure*}

\section{Discussion} \label{sec:discussion}

\subsection{Local Stellar Halo Luminosity Function and Number Density in the Literature} \label{sec:lit}

\begin{table*}
\begin{center}
  \caption{Estimates\tablenotemark{a} of $\rho_\mathrm{halo}$ --- the local stellar halo number density.}
\label{table:lit}
\begin{tabular}{c|cccccc}
\hline
Authors & Method\tablenotemark{b} & magnitude range & $\rho_\mathrm{halo}$ & Notes & \\
       &        & mag   & stars pc$^{-3}$         &       & \\
\hline
\citet{SchmidtMaarten1975} & $V_\mathrm{t}>250\kms$ &  $1 < M_\mathrm{pg} < 13$  &   $1.9\times10^{-4}$ &  &  \\
\hline
\citet{Fenkart1977} & three-color photometric & $3 < M_G < 8$     &  $\approx 5\times 10^{-4}$  &  $G$-band on & \\
 & method using $RGU$ & & &   the Basel system. & \\
 & due to $UV$ excess & & & estimated from & \\
 &                    & & & table 2 and fig 2\tablenotemark{c}.                 & \\

\hline
\citet{Eggen1979halolf} & $\mu>1''$\,yr$^{-1}$ & $6 < M_V < 14$   & $ 5 \times10^{-4} $ &  summed LF in table 6\tablenotemark{c} & \\
 & eccentricity$>0.42$ &   &  &   & \\

\hline
\citet{Eggen1983} & $\mu>0''.7$\,yr$^{-1}$ & $4 < M_V < 14$   & $ 2 \times10^{-3} $ &  & \\ 
                  & [Fe/H]$<-0.6$         &            &                     &  & \\
                  & eccentricity$>0.42$   &            &                     &  & \\

\hline
\citet{Chiu1980-III} & reduced proper motion & $4 < M_V < 12$   &  $6 \times 10^{-4}$    & $\rho$ extrapolated by 30\% & \\
\hline
\citet{Hartwick1983} & metal poor subdwarfs & $4.8 < M_V < 6.0$   & $9\times10^{-6}$  & \\
\hline
\citet{Hartwick1983} & metal poor giants & $-1.8 < M_V < 1.4$& $4 \times 10^{-7}$& summed LF in table 3\tablenotemark{c}  & \\
\hline
\citet{Lee_Sang-Gak1985} & metallicity & $4 < M_V < 15$ & $\approx 2 \times 10^{-3}$& summed LF from table 2\tablenotemark{c}& \\

\hline
\citet{Bahcall1986} & $V_\mathrm{t}>220\kms$ & $4 < M_V < 11$ & $(9.5\pm1.3) \times 10^{-5}$  &    & \\
\hline
\citet{Dawson1986} & reduced proper motion & $ 3 < M_V < 14$&  $1.3 \times 10^{-4}$  &  & \\
 & $\mu > 1''$ yr$^{-1}$, parallaxes & &   &  & \\

\hline
\citet{Lee_Sang-Gak1991} & reduced proper motion &  $4 < M_V < 12$  &  $4.0 \times 10^{-4}$  & summed LF table 2\tablenotemark{c} & \\
\hline
\citet{Richer1992} & faint ground-based star& $5 < M_V < 14$ &  $\approx 3 \times 10^{-4}$ & obtained from fig 2(c)\tablenotemark{c}   & \\
& counts at $b=73^\circ$ &  &   &     & \\

\hline
\citet{Lee_Sang-Gak1993} & reduced proper motion &  $4 < M_V < 10$  &  $1.4 \times 10^{-4}$ & summed LF table 3\tablenotemark{c}  & \\

\hline

\citet{Dahn1995} & $ 0''.8 < \mu < 2''.5  \, \mathrm{yr}^{-1}$ &$  5 < M_V < 15 $ &  $2.8 \times 10^{-4}$ & CCD-based parallaxes& \\

 & $V_\mathrm{t}>220\kms$ &    &  & & & \\

\hline
\citet{Lee_Sang-Gak1995} & $\mu > 0''.2$ yr$^{-1}$ & $4 < M_V < 8$  &  $2.3 \times 10^{-5}$ &  & \\
& $V_\mathrm{t}>220\kms$ & &  &  & \\

\hline
\citet{Fuchs1998} & Nearby Star Catalog  &  $6 < M_V < 14$  &  $2.3 \times 10^{-4}$  & & \\
\hline
\citet{Gould1998} & faint HST star counts &  $6.5 < M_V < 14.5$  & $1.0 \times 10^{-4}$   & summed LF from fig 3 &\\
 & $|Z| > 8 $ kpc &    &   &  & \\
\hline
 \citet{Gizis1999}   & $0''.1 < \mu<0''.375 \, \mathrm{yr}^{-1}$ & $11 < M_V < 13$  &  $7.0 \times 10^{-5}$  & summed LF from fig 6 & \\
 & reduced proper motion & &  & & \\
 & $V_\mathrm{t} > 200$ km\,s$^{-1}$  & & &  & \\
\hline
\citet{Cooke2000} & $\mu>0''.2 \, \mathrm{yr}^{-1}$  &  $4 < M_V < 14$ &  $2.7 \times 10^{-4}$  & summed LF from fig 6\tablenotemark{c}& \\
 & reduced proper motion & & &   & \\
\hline
\citet{Gould2003,Gould2004erratum2003} & $\mu>0.2 '' \, \mathrm{yr}^{-1}$ &  $3 < M_V < 15$   & $2.2 \times 10^{-4}$&  summed LF from fig 2\tablenotemark{c}& \\
 & reduced proper motion &     &    & \\
\hline
\citet{Digby2003} & $ 40.5 < \mu < 160\, \mathrm{mas\, yr}^{-1}$ &  $4 < M_V < 13$  &  $1.8\times10^{-4} $   &  summed LF from fig 24\tablenotemark{c} & \\
 & $V_\mathrm{t}> 200 \kms$ & & &  & \\

 & reduced proper motion & & &  & \\
\hline
\citet{Rojo2003} & metallicities & $6 < M_V < 16$    & $4.0 \times 10^{-4}$   &  summed LF from fig 10\tablenotemark{c} & \\
 & $(U,V,W)$ velocities& & &  & \\
\hline
This study & $V_\mathrm{t} > 250 \kms$ & $-3 < M_V < 13$    & $1.7 \times 10^{-4}$   &  summed LF from Fig. \ref{fig:lf-lit} & \\
 & & & & of this paper & \\

\hline
\end{tabular}
\end{center}
\tablenotetext{a}{Estimates of $\rho_\mathrm{halo}$ within the given magnitude range. This is the local stellar halo number density at $Z=0$\, pc.}
\tablenotetext{b}{Dominant method used to select halo sample.}
\tablenotetext{c}{Note on estimation of $\rho_\mathrm{halo}$ using the figure (fig) or table from the referenced author(s).}
\end{table*}

Before detailing the comparison of our stellar halo LF with previously measured LFs in Fig. \ref{fig:lf-lit}, we provide a historical perspective on the many and varied measurements of the halo LF and local number density. We provide a comparison of the local stellar halo number density estimates from the literature in comparison with our own work in Table \ref{table:lit}, including the method and magnitude range used by the authors in determining the number density.

The key ingredients to the current study are the parallaxes and proper motions from the all-sky survey {\it Gaia}. Taking into account the stellar halo population's characteristically large velocity dispersion and low-rotation, these together allow us to separate halo stars in a nearby sample by selecting stars with high tangential velocities.

Sky surveys collecting proper motions have a beginning in the early 1800's with the work of Friedrich Argelander at the Old Observatory in Turku, Finland and his published {\AA}bo Catalog \citep{Argelander1835}. The reader is also referred to historical overviews of the very early periods by, {\it e.g.,} \citet[][]{Gefwert1975,Mattila_Kalevi2004}.

The first measurements of the disk LF appeared in the early 1900's with the work of, {\it e.g.}, \citet{Kapteyn1902,Luyten1938,Kuiper1942}. 

During this time the concept of the stellar halo and stellar populations themselves were being established, {\it e.g.}, see the review paper by \citet{Sandage1986}. It was not until \citet{SchmidtMaarten1975}, that the first stellar halo LF was measured; he used high proper motion stars with trigonometric parallaxes (measured for the majority of the sample) to select halo stars with $V_\mathrm{t}>250\kms$ and obtained a local stellar halo number density of $1.9\times10^{-4}$ stars pc$^{-3}$. 

\citet{Fenkart1977} presented the LF for halo stars selected by a purely photometric procedure to statistically separate disk and halo stars using `$UV$ excess' as a metallicity indicator. The LF covers a quite narrow range of absolute magnitude from $3 < M_G < 8$, where $G$ here refers to the Basel photometric system. To derive the local number density of halo stars, \citet{Fenkart1977} extrapolates the LF down to $M_G \approx 16$. \citet{Fenkart1977} found ratios of the disk-to-halo star local mass densities (at the Milky Way mid-plane) in the range 4:1 to 50:1, with results being quite model dependent. 

\citet{Eggen1979halolf,Eggen1983} studied high proper motions stars ($>1$ arcsec\,yr$^{-1}$, and $>0.7$ arcsec\,yr$^{-1}$ respectively), deriving space motions (calculated from the parallaxes, radial velocities, and proper motions) and orbital eccentricities. \citet{Eggen1979halolf} found a local stellar halo number density of $5 \times 10^{-4}$ stars pc$^{-3}$, where halo stars are defined as having highly eccentric ($e > 0.42$) orbits. \citet{Eggen1983} analyzed a larger sample, separating stars into disk and halo types, finding a halo LF that rises sharply, showing an excess of faint stars, down to $M_V \approx 14$. \citet{Eggen1983} derived a space number density for halo stars of $2\times10^{-3}$ stars pc$^{-3}$, significantly higher than the \citet{Eggen1979halolf} study. \citet{Eggen1983} found that halo stars comprise a rather high fraction (1.5\%) of local stars, by stellar mass density (including main sequence stars and white dwarfs).

\citet{Chiu1980-III} used a complete set of accurate proper motions in four of Kapteyn's `Selected Areas' to very deep (photographic) magnitudes ($V < 20.5$) \citep[SA 51, SA 57, and SA 68, see details of observations in][]{Chiu1977-I,Chiu1980-II} on the Lick 3\,m, Hale 5\,m and KPNO 4\,m telescopes. This technique gives access to good numbers of halo stars high above the disk, and using reduced proper motion analysis and colors, kinematics could be probed and compared to the solar neighborhood. This pioneering technique became considerably more effective after wide-field electronic detectors came into use, as the photographic colors in only two bands hampered the ability to analyze the stars by chemistry. \citet{Chiu1980-III} derived the halo LF in the range $4 < M_V < 12$ and estimated a local mass density for the halo of $3 \times 10^{-4}$ M$_\odot$ pc$^{-3}$, where they included a 30\% correction for stars with $M_V < 4$ and $M_V>12$. 

It is worth noting that \citet{Bahcall1983} used the halo LF to make the first estimates of the halo's total luminosity and mass.

\citet{Hartwick1983} analyzed samples of dwarf and giant stars with metallicity estimates, to measure the local density of the halo and its LF, using the `V/Vmax' method. Extremely metal-deficient giants came from the catalog of \citet{Bond1980} and halo dwarfs from the catalog of \citet{Carney1979ApJ.233.877}. The LF was derived in the $V$ band, rising rapidly in the range $-1.6 < M_V< 5.8$. This was an influential study in the use of such stars to constrain the formation and early evolution of the Galaxy. \citet{Hartwick1983} found a space number density for local halo dwarfs in the narrow absolute magnitude range $4.8 < M_V < 6.0$ of $9 \times 10^{-6}$ stars pc$^{-3}$.
We have integrated the LF for halo giants in \citet{Hartwick1983} (their table 3) in the absolute magnitude range $-1.8 < M_V < 1.4$, obtaining a 
space number density of $4 \times 10^{-7}$ stars pc$^{-3}$, after correcting by a factor of two as the source catalog for giants is confined to extremely metal-poor stars of the halo (discussed in more detail below in Sec \ref{sec:litcompare}). 

\citet{Reid1984}, \citet{Bahcall1986}, and \citet{Dawson1986} pioneered the use of computer simulations to aid in measuring the halo LF, particularly for deep star count samples with proper motions, and/or high velocity stars. These techniques are seminal for the analysis in this paper.

\citet{Lee_Sang-Gak1985} used kinematically unbiased samples of halo stars (i.e. selected by metallicity only) to estimate the local mass density of the halo, obtaining $6.3 \times 10^{-4}$ M$_\odot$ pc$^{-3}$. This did not differ substantially from estimates derived from kinematically biased samples, indicating the the kinematic effects could be corrected out. \citet{Lee_Sang-Gak1991} used reduced proper motions to select halo subdwarfs, deriving a remarkably flat halo LF in the range $4 < M_V < 12$, and a disk to halo ratio by space density of approximately 100:1. 

In a seminal work, \citet{Richer1992} used deep imaging at high Galactic latitude in small fields with electronic detectors in $V$, $R$ and $I$ on the Canada-France-Hawaii Telescope to derive the halo LF in the range $5 < M_V < 14$ and mass function, both sharply rising.
Estimates of the mass of the stellar halo in \citet{Richer1992} and \citep{Lee_Sang-Gak1993} showed that this Milky Way component was far too light to account for flat rotation curves, topical at the time. 

\citet{Dahn1995} used a proper motion selected sample of 144 subdwarfs (within distances $\le300$\,pc from the Sun) from the Luyten Half-Second Catalogue \citep{Luyten1979} and pioneered the use of CCD detectors for parallax measurements from the ground at the U.S. Naval Observatory (USNO) \citep{Monet1992}. This work produced superb CMDs of the lower main sequence of the halo and a halo LF in the range $ 8 <  M_V < 15$, showing a clear peak $M_V \approx 12$ and a decline in the star counts down to $M_V \approx 14$. This turn over in the LF is similar to the well determined LF for the disk stars. They obtained a local halo space number density of $2.8 \times 10^{-4}$ stars pc$^{-3}$ by combining their LF in the range $ 8 <  M_V < 15$ with the LF from \citet{Bahcall1986} in the range $5 <  M_V < 8$.

Using the same method as this paper, \citet{Lee_Sang-Gak1995} used a tangential velocity cut of $>220\kms$ on a sample of 233 high proper motion stars ($\mu > 0.2$ arcsec\,yr$^{-1}$) to measure the bright end ($4 < M_V < 8$) of the Galactic halo subdwarf LF. Combined with work by \citet{Bahcall1986, Dawson1986} and \citet{Dahn1995}, the LF shows a rise to a stellar number density of $10^{-4}$ stars mag$^{-1}$ pc$^{-3}$ and the turn over beyond $M_V \approx 12$. 

\citet{Reid1996} combined both ground-based data from the Keck 10\,m telescope and fainter data from the {\it HST} Hubble Deep Field \citep{Williams1996} to build the halo LF through deep star counts. Combined with halo LF modeling, they caution that the dim end of ground-based measurements of the stellar halo LF may be contaminated by disk stars which can lead to an erroneous increasing LF at $M_V>10$ mag \citep[{\it e.g.},][]{Richer1992}.

\citet{Fuchs1998} used CNS4 \citep{Jahreiss1997} with improved parallax and proper motion measurements from Hipparcos to measure the stellar halo LF from 15 subdwarfs within 25 pc, which lie in the absolute magnitude range $6 < M_V < 14$,
which yields a space number density of $2.3 \times 10^{-4}$ stars pc$^{-3}$. These were carefully selected based on the characteristics of subdwarfs, checking that they lied along the subdwarf branch in the color-magnitude diagram and had high tangential velocities and low metallicities. They derived a firm lower limit to the local mass density for the stellar halo of $1 \times 10^{-4}$ M$_\odot$ pc$^{-3}$. 

\citet{Gould1998} measured the halo LF using 166 halo subdwarfs with inferred distances of $>8$\,kpc above the disk, located in 
53 very deep ($V \approx 23-26$ mag) fields (one of which was the Hubble Deep Field) with a total area of 221 arcmin$^2$ taken with WFPC2 on {\it HST}. Their LF was measured in the range $ 6.5 < M_V < 14.5$. Their LF was flatter and at a lower amplitude than LF determinations using star samples much closer to the Sun, the cause for which was unclear. They derived a local mass density for the stellar halo (including unseen companions and stellar remnants) of $6.4 \times 10^{-5}$ M$_\odot$ pc$^{-3}$. 

\citet{Gizis1999} used a proper motion and color selection for halo M dwarfs from plate scans from the Palomar Observatory Sky Survey to measure the halo LF in the range $11 < M_V < 13$. This LF is broadly consistent with others at the time \citep{SchmidtMaarten1975, Bahcall1986, Dahn1995}, all of which have a substantially higher amplitude, as just noted, than the HST counts of \citet{Gould1998}. 

By adopting kinematic models for the disc and halo populations \citet{Cooke2000} derived the halo LF in the range $4 < M_V < 14$ from a proper motion-based ($\mu > 0.2$ arcsec\,yr$^{-1}$) selection of halo M dwarfs with $BVRI$ photometry, found from plate scans of Kapteyn's Selected Area 94. The kinematic modeling technique to derive completeness levels is similar to the one adopted in this paper. 

\citet{Gould2003,Gould2004erratum2003} used reduced proper motions to select 4564 subdwarfs from the revised New Luyten Two-Tenths (NLTT) Catalogue \citep{GouldSalim2003,Salim2003} and from this sample estimated the LF using a maximum likelihood method. They derived an LF in the range $3 < M_V < 15$, with a clear peak at $M_V \approx 11$ and a local number density of approximately $2.2 \times 10^{-4}$ stars pc$^{-3}$.

\citet{Digby2003} used a reduced proper motion sample from plate scans of the Palomar Observatory Sky Survey matched with SDSS to measure the halo LF using $\sim 2600$ subdwarfs in the range $4 < M_V < 13$, for $V_\mathrm{t} > 200\kms$ showing an LF with a clear peak at $M_V \approx 11$.

\citet{Rojo2003} used the medium-resolution spectra and proper motions from the Cal\'{a}n-ESO survey \citep{Ruiz2001} to select 18 halo subdwarfs based on spectral index/metallicity and $V$ velocity (i.e. the velocity in the direction of rotation of the disk in a Cartesian Galactic reference frame). Adopting $V<-100 \kms$, they measured the LF in the range $6 < M_V < 16$, and found a rising LF all the way to faint end, rather than a turn over, although the sample size is small.

Since the beginning of the 21st century, work on the stellar halo LF has remained dormant.

\subsection{Comparison of Gaia DR3 LF with Previous Results} \label{sec:litcompare}

Besides the current study (black markers in Fig. \ref{fig:lf-lit}), the most recent halo LF is a careful compilation of several LFs \citep{Casertano1990,Gizis1999,Dahn1995,Gould2003,Digby2003} reviewed by \citet{Reid2005}. 
Since \citet{Hartwick1983} give the number density of stars per every 0.4 $V$-band mag bin, we multiply the number density by (1/0.4) to convert to number density per magnitude. \citet{Hartwick1983} sourced the halo giant sample from the extreme metal-deficient catalog of \citet{Bond1980}. Taking into account the metallicity distribution of the inner Galactic stellar halo \citep[{\it e.g.},][]{Youakim2020}, we correspondingly plot the \citet{Hartwick1983} halo giant LF increased by a factor of two. We update \citet{SchmidtMaarten1975} using $V$-band photometry from Simbad \citep{Wenger2000}. For \citet{SchmidtMaarten1975,Dahn1995,Lee_Sang-Gak1995,Fuchs1998,Rojo2003} we plot Poisson uncertainties \citep{Gehrels1986}.

\citet{Gould2003} confirmed the bump in the LF found by \citet{Dahn1995} at $M_V\approx 11$. As the uncertainties increase towards the last few bins we are unable to confirm a turn over in our current study of the stellar halo LF, although we are able to confirm that within the magnitude range probed within our sample we see the maximum of the bump feature reached at similar magnitudes ($M_V\sim10-12$).

Our value for the local stellar halo luminosity density measured from $M_V$, $1.5\times10^{-5}$\,L$_\odot$\,pc$^{-3}$, falls within the range estimated in $V$-band by the early studies of \citet{Bahcall1980paperI,Bahcall1983}. The two-component standard Galactic model of \citet{Bahcall1980paperI} gives a local halo luminosity density of $7.7\times10^{-5}$\,L$_\odot$\,pc$^{-3}$. \citet{Bahcall1983} estimate the local stellar halo $V$-band luminosity density, using star counts, finding $(2-8)\times10^{-5}$\,L$_\odot$\,pc$^{-3}$ and using high velocity stars, finding $(3-13)\times10^{-5}$\,L$_\odot$\,pc$^{-3}$.

We find a total $V$-band luminosity for the Milky Way stellar halo (integrating over a broken power-law distribution to a Galactocentric distance of 100\,kpc) of $4.1\times10^8$\,L$_\odot$ and absolute magnitude of $-16.7$ mag in the $V$-band.

The early Milky Way model of \citet{de_Vaucouleurs1978} consists of two components, the disk and spheroid (this component represents the combined bulge and stellar halo). Using a spherical halo component, they found a $V$-band total spheroid luminosity of $5.3\times10^9$\,L$_\odot$ and absolute magnitude of $-19.54$ mag excluding dust; and $4.5\times10^9$\,L$_\odot$ and $-19.36$ mag dimmed by dust (as an outsider's face-on view of the Milky Way). Using a flattened spheroid component with ellipsoidal axis ratio of 0.6, \citet{de_Vaucouleurs1978} found a $V$-band total spheroid luminosity of $6.85\times10^9$\,L$_\odot$ and absolute magnitude of $-19.82$ mag (after extinction corrections).

The standard model of \citet{Bahcall1980paperI} gives a total luminosity of the Milky Way stellar halo of $1.9\times10^9$\,L$_\odot$.
\citet{Deason2019.490} used star count data of red giants from {\it Gaia} DR2, finding a total Milky Way stellar halo luminosity of $7.9 \pm 2.0 \times 10^8$\,L$_\odot$ excluding
the satellite Sagittarius and $9.4 \pm 2.4 \times 10^8$\,L$_\odot$ including Sagittarius. The difference ratio is 0.6 when we compare our total Milky Way stellar halo luminosity in $G$-band of $4.6\times10^8$\,L$_\odot$ with the estimate of \citet{Deason2019.490} excluding Sagittarius.

\subsection{Features in the Luminosity Function} \label{sec:lffeatures}

The general shape of the nearby stellar halo is remarkably similar to that of the disk. 
We make note of several features in the nearby stellar halo LF that are comparable to features discussed in the literature that are seen in the disk LF as well as cluster LFs.

The horizontal branch stars are clearly visible in the giant LFs: the red clump appearing at $M_G=0.4$ \citep{GaiaCollaborationSmart2021} and the blue horizontal branch stars near $M_G\sim0.7$.

The LF of the nearby {\it Gaia} disk sample peaks at $M_G=10.5$ \citep{GaiaCollaborationSmart2021}. Our {\it Gaia} halo LF reaches maximum at similar magnitude. Our halo LF remains quite flat as it reaches dimmer magnitudes and the uncertainties increase, thus detecting a credible turn over in our measured halo LF becomes difficult. Prospects for decreasing these errors include collecting dim subdwarfs from other telescopes such as {\it HST}, {\it CSST}, {\it JWST}. The disk LF, on the other hand, clearly turns over after reaching the maximum; {\it e.g.}, using the high quality, volume-complete {\it Gaia} Catalog of Nearby Stars (distances within 100 pc), \citep{GaiaCollaborationSmart2021} show that after reaching the maximum, the disk LF continues to fall down to absolute magnitudes $M_G\sim16-17$.

The flattening of the solar neighborhood disk LF around $M_V\sim5-9$ was noted early on \citep[{\it e.g.},][]{Arakelyan1968,Mazzitelli1972,Wielen1974,Wielen1983,Upgren1981}, and is often called the Wielen Dip. This feature has also been noted as detected in open cluster LFs \citep[{\it e.g.},][]{Mazzitelli1972b,Lee_See-Woo1995,Guo_Difeng2021}. We observe this flattening feature in the nearby stellar halo LF also near $M_V\sim5-9$, similar to previous detections of the disk and cluster data. This feature was also noted in the stellar halo LF determined by \citet{Bahcall1986,Lee_Sang-Gak1995} and included in the modelled stellar halo LF of \citet{Bahcall1984ApJS.55.67}. \citet{D-Antona1998} elaborated upon this feature using metal-poor Population II theoretical models. By comparing the LFs from previously measured Galactic stellar halo samples \citep[{\it e.g.},][]{Gould2003,Digby2003}, we can see evidence of the Wielen Dip, although these authors have not specifically pointed out or discussed the feature.

Direct comparison of the stellar halo LF to globular clusters is hampered by the evolution of the cluster such as through relaxation and mass segregation within the cluster and the loss of stars due to the dynamics of the cluster as it orbits the Galaxy \citep[{\it e.g.},][]{Krumholz2019}. 
We note that the common features such as the peak of the LF located at low luminosities and the flattening of the LF in the $M_V\sim5-9$ range are commonly found in the present-day LFs of Galactic globular clusters \citep[{\it e.g.},][and references therein, although these authors do not always discuss such features]{Paust2010}.

The luminosity function and the stellar mass function are related by the mass-luminosity relation. Features in the luminosity function can thus reflect features in the mass function or in the mass-luminosity relation determined by the physics of the stars themselves, as emphasized by, {\it e.g.}, \citet{D-Antona1983} and \citet{Kroupa1990}.
For low mass stars ($<1M_\odot$), the differences in the physics of the star depending on their mass due to, {\it i.e.}, ionization of hydrogen, dissociation/recombination of molecular hydrogen H$_2$, electron degeneracy, leads to detectable features in such relations as the CMD, HR diagram, luminosity function, and mass-luminosity function \citep[see, for example,][]{Copeland1970,Kroupa1990,D-Antona1996,Kroupa2002,Salaris2005}.
Observing features in the stellar halo LF is significant as this allows comparison of the stellar physics of 
metal-poor halo stars to other populations, {\it i.e.}, disk, clusters, etc.

Although this is the largest yet sample of halo stars used to measure the LF, the very fine feature detected by \citet{Jao2018} in the disk may not yet be easily detected in the halo data. A more detailed search for the Jao Gap/M-dwarf Gap in the stellar halo is left for future efforts.

\section{Conclusions and Future Work} \label{sec:conclusion}

We measure the stellar halo luminosity function using a magnitude complete, distance limited, tangential velocity-selected sample of nearby stars in {\it Gaia} DR3. 
The stellar halo sample contains more stars and probes a wider range of absolute magnitudes than all previous studies.

We convert the {\it Gaia} $G$ band to Johnson-Kron-Cousins $V$ and find the stellar halo luminosity compares well with previous results.
In further efforts we will use this sample to convert the stellar halo luminosity function to other commonly used photometric systems for convenient use for models and observational analyses.
The LF is one of the first steps to measure the mass function of stars. Only recently has the mass of a subdwarf been measured \citep{Rebassa-Mansergas2019}.
We will use low metallicity theoretical models along with this result to measure the stellar halo mass function and mass-to-light ratio.

\acknowledgments
Thanks to Dan-Wei Fan and the China-Virtual Observatory of the National Astronomical Data Center for aiding with data access. Thanks to Mauri Valtonen for discussions on the early history of proper motion studies. This research has been supported by the National Natural Science Foundation of China under grant number 12303023. 
This work has made use of data from the European Space Agency (ESA) mission
{\it Gaia} (\url{https://www.cosmos.esa.int/gaia}), processed by the {\it Gaia}
Data Processing and Analysis Consortium (DPAC,
\url{https://www.cosmos.esa.int/web/gaia/dpac/consortium}). Funding for the DPAC
has been provided by national institutions, in particular the institutions
participating in the {\it Gaia} Multilateral Agreement.
This research has made use of the SIMBAD database, operated at CDS, Strasbourg, France.

\software{{\tt Astropy} \citep[v5.1;][]{astropy2013,astropy2018}, 
  {\tt galpy} \citep[v1.9.0;][]{Bovy2015}, {\tt matplotlib.pyplot} \citep[v3.7.0;][]{Hunter2007}, {\tt NumPy} \citep[v1.23.5;][]{Oliphant2006,Walt2011,Oliphant2015,Harris2020array}, {\tt pandas} \citep[1.5.3;][]{pandas2022}, {\tt SciPy} \citep[v1.10.0;][]{Virtanen2020}, {\tt TOPCAT} \citep[v4.8-8;][]{Taylor2005}.
}

{\it Data Availability Statement:}\\
The data presented in the figures are available upon request from the
authors.

\bibliographystyle{apj} 
\bibliography{Bird_2026_lf}




\appendix

\section{Initial {\it Gaia} DR3 Data Query}
\label{app:adql}

We use Astronomical Data Query Language \citep{Osuna2008} with the following query to perform an initial selection of data within a sphere of 1\,kpc radius about the Sun, yielding $39\,669\,995$ sources from {\it Gaia} DR3 \citep{GaiaCollaborationPrusti2016, GaiaCollaborationVallenari2023, Babusiaux2023}\footnote{Descriptions for {\it Gaia} DR3 columns are found at \url{https://gea.esac.esa.int/archive/documentation/GDR3/Gaia_archive/chap_datamodel/sec_dm_main_source_catalogue/ssec_dm_gaia_source.html}.},\\
{\tt
SELECT * FROM gaiadr3.gaia\_source\\ WHERE visibility\_periods\_used>5\\ AND astrometric\_excess\_noise<0.5\\ AND parallax > 1\\ AND parallax\_over\_error > 5\\ AND phot\_bp\_mean\_flux\_over\_error>20\\ AND phot\_rp\_mean\_flux\_over\_error>20\\ AND phot\_g\_mean\_flux\_over\_error>20\\ AND phot\_bp\_rp\_excess\_factor \\< 1.2*(1.2+0.03*\\ (phot\_bp\_mean\_mag-phot\_rp\_mean\_mag)*\\ (phot\_bp\_mean\_mag-phot\_rp\_mean\_mag)).
}\\
Further selection criteria applied to this initial sample are fully described in Sect.\,\ref{sec:data}.

\section{Conversion of {\it Gaia} photometry to Johnson-Kron-Cousins $V$ band}
\label{app:GV}

\citet{GaiaCollaborationMontegriffo2023} provide synthetic photometry in multiple bands for {\it Gaia} data. We match our halo star sample after removing white dwarfs, to the synthetic $V$ data. Not all {\it Gaia} data have synthetic photometry estimates. We use the matches to estimate the relation between {\it Gaia} $G$ and BP$-$RP and Johnson-Kron-Cousins $V$.

\begin{figure}
\includegraphics[width=0.5\columnwidth]{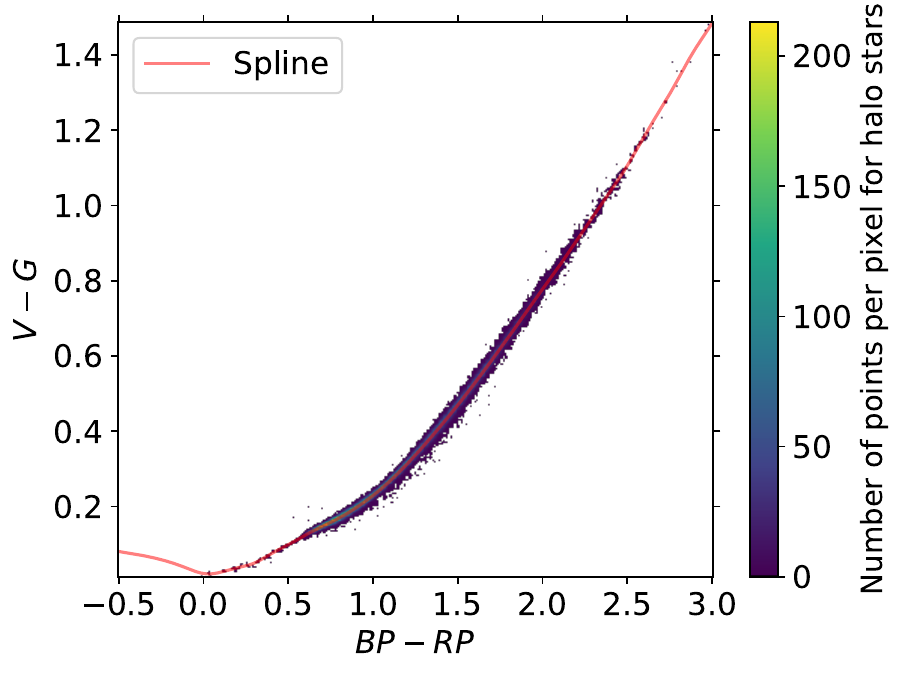}
\caption{Spline fit (red line) of the halo data with synthetic $V$ (scatter-density map). Used to convert $G$ to $V$ for full {\it Gaia} stellar halo dataset. The color bar indicates the number counts.
} 
\label{fig:GVconvert}
\end{figure}

Fig. \ref{fig:GVconvert} shows the relation that we use to estimate $V$ from {\it Gaia} $G$ and BP$-$RP for our halo stars (excluding white dwarfs). We use the spline function of {\tt scipy.interpolate.make\_smoothing\_spline} to estimate the relation. Using this relation we make an estimate of $V$ for all halo stars in our sample used for measuring the LF. When using the $V$ magnitudes to estimate the stellar halo LF, we set the limiting magnitude to $V=18$.

\section{Determination of halo $V_\mathrm{t}>250\kms$ cut and corresponding LF correction factor}
\label{app:mwmodel}

\subsection{Milky Way model}

\begin{figure*}
\includegraphics[width=1\columnwidth]{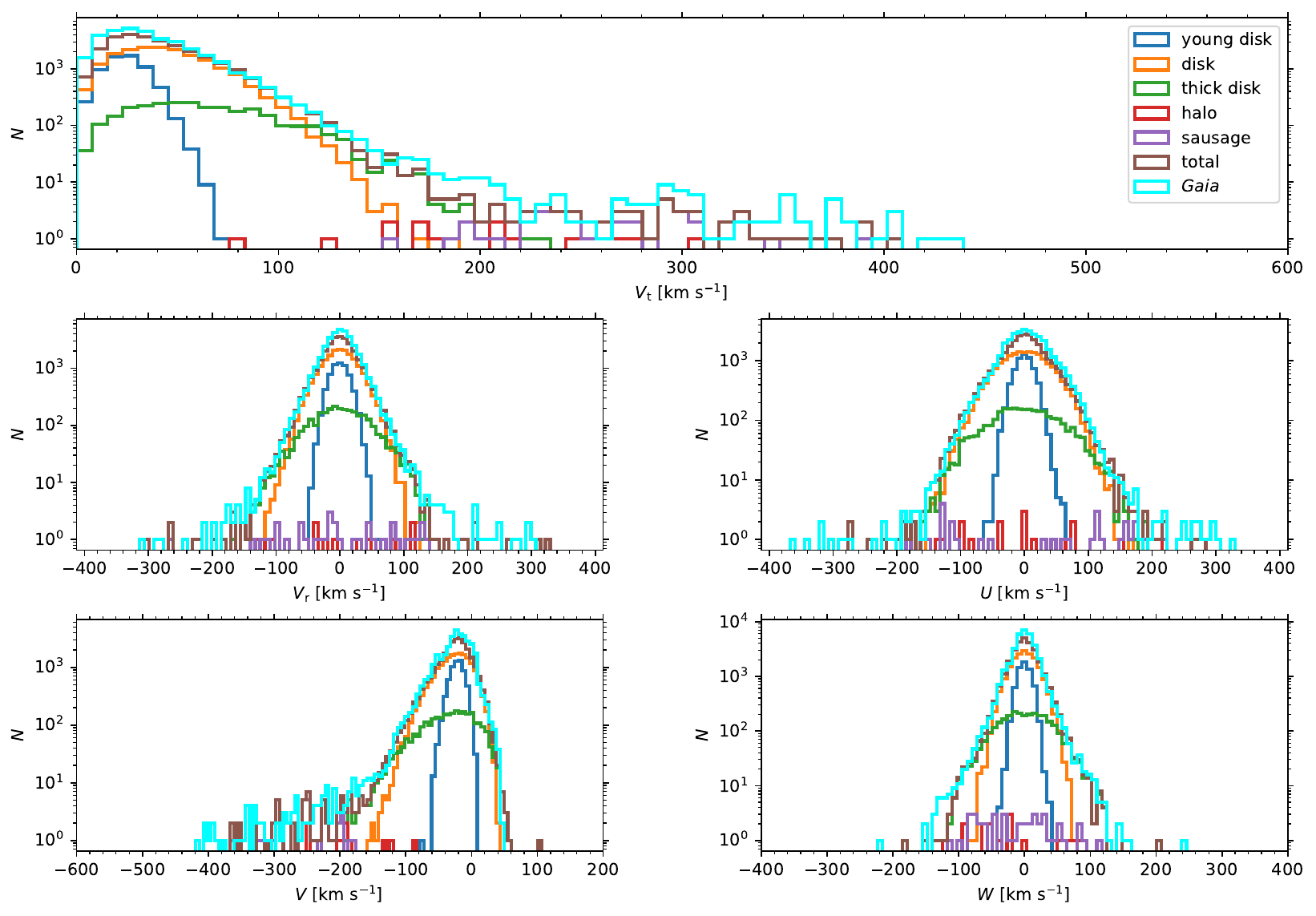}
\caption{Milky Way model (young disk-blue, thin disk-orange, thick disk-green, underlying halo-red, $Gaia$-Enceladus-Sausage satellite-purple, and all components in total-brown) and our selected $Gaia$ sample within a sphere of radius 1 kpc with radial velocity measurements (cyan). Velocity components shown for tangential and radial velocity $V_\mathrm{t}$ (uppermost) and $V_\mathrm{r}$ (middle left), respectively, and the 3D Cartesian velocity components $(U,V,W)$ (middle right, lowermost left, lowermost right, respectively).
}
\label{fig:mwmodel}
\end{figure*}

We plot the velocity components of our {\it Gaia} sample using the cyan curve in Fig. \ref{fig:mwmodel}. These stars consist of those selected in Sect. \ref{sec:dataselect} that have radial velocities measured by {\it Gaia}. The uppermost panel plots the $V_\mathrm{t}$ distribution, from which we can see the dominating peak of disk stars and the highest velocity stars belonging to the stellar halo. Note that the vertical axes are in logarithmic scale. Our aim is to fit this distribution with the five component model sufficiently well, that we can select an effective cutoff velocity for a very high purity halo star sample, while also yielding a confident estimate of the missing halo fraction. 

In the following we describe our Milky Way model and its five components. The model is summarized in Table \ref{table:mwmodel} and we plot the total model velocity distribution (brown) in Fig. \ref{fig:mwmodel} as well as each model component (see figure).

In this model, we set the total local stellar density at the Sun as 0.081 stars pc$^{-1}$ \citep[{\it e.g.},][]{GaiaCollaborationSmart2021,Reid2002}.

Each component is modeled by (1) local density at $Z=0$, (2) a velocity ellipsoid in the cardinal directions and a mean rotational velocity around the Galactic center and (3) a scale-height vertically out of the Galactic disk. For the halo, the adopted velocity distributions are Gaussian in all three components. For the disk populations, the radial and vertical velocity distributions are Gaussian, but the rotational velocity distribution is asymmetric, and follows the precepts of \citet{Dehnen1999} for disk populations based on energy and angular momentum.

To determine the number of stars for each component within our model, we take into account that the density of each population falls with increasing height from the Galactic mid-plane. We use $\mathrm{sech}^2(Z/(2\,Z_h))$ to represent the density falloff where for the young disk, thin disk, thick disk, halo, and {\it Gaia}-Enceladus-Sausage, we use scale heights $Z_h$ of 76, 320, 800, 6000, 6000 pc, respectively. The scale heights reflect the values found in the literature for the thin and thick disk and halo as reviewed by \citet{Bland-Hawthorn2016} and for the young disk, {\it e.g.}, \citet{Reed2000,Kong2008,Yu2021,Quintana2025}. The stellar halo near the Sun has a radial density profile that falls as a power law of $\sim -2.5$ \citep[{\it e.g.},][]{Bland-Hawthorn2016}. Within our one kiloparsec sample, this is well estimated by the $\mathrm{sech}^2$ density law with a scale height of 6000 pc. As we use the scale height to correctly model the number of halo stars in our volume, we note that the total number of stars in the volume with the radial scale factor model is similar to the total number of stars when using the more realistic radial power law density model. This is to say, the sum of halo stars in our sample towards the Galactic Center added with to those away from the Galactic Center is similar to the sum of using $\mathrm{sech}^2$.

We have set the fractional contribution of each component to the model as 0.250, 0.650, 0.100, 0.00518, and 0.00155 for the young disk, thin disk, thick disk, underlying halo, and {\it Gaia}-Enceladus-Sausage, respectively. The halo itself is composed of a combination of a fractional contribution of 0.25 of the remaining halo stars after the fractional contribution $1-0.25$ due to the {\it Gaia}-Enceladus-Sausage satellite, making a total halo fraction equal to 0.00207 of all stars in the model.
The fraction of {\it Gaia}-Enceladus-Sausage is compatible with the work by \citet{Belokurov2018,Necib2019_874,Lancaster2019,Iorio2021,Wu2022}.

A simple representation of the disk's velocity distributions is to use normal, or Gaussian distributions. The very high quality {\it Gaia} data show that this is a far too simple model for the rotational component of the 3D velocity ellipsoid of the disk. We can see this in the $V$-component of the velocity (lowermost left panel of Fig. \ref{fig:mwmodel}). The curve is highly skewed toward smaller $V$-values for which a single Gaussian cannot account. Because of the asymmetric drift we make use of the {\tt galpy} \citep{Bovy2015} implementation of the warm disk model of \citet{Dehnen1999}. As we can see in Fig. \ref{fig:mwmodel}, the \citet{Dehnen1999} distribution function is a very good representation the observed skewness of the smaller $V$-velocities of the data. We use this model for all three disk components. The mean and dispersion for each $(U,V,W)$ cartesian velocity component for the thin, young, and thick disks are recorded in Table \ref{table:mwmodel}. This introduces one additional parameter into the fitting, which is the scale length $h_R$ of the stellar density in the disk in the radial direction. In our fitting, we adopt $h_R = 4$ kpc. 

The halo and {\it Gaia}-Enceladus-Sausage models and parameters are chosen for consistency with the results presented by \citet{Belokurov2018,Necib2019_874,Lancaster2019,Iorio2021,Bird2021,Wu2022}. The halo is modeled by a single Gaussian and the {\it Gaia}-Enceladus-Sausage is modeled by a double-Gaussian.

The model for the underlying halo component has mean velocities $(0,-220,0)\kms$ and a velocity ellipsoid $(140,100,100)\kms$. The velocity of the {\it Gaia}-Enceladus-Sausage is modeled by two Gaussian distributions with equal but opposite mean $U$ velocity components, such that the means are $(\pm150,-240,-10)\kms$ and dispersions $(90,40,60)\kms$.

\begin{table*}
  \caption{Milky Way model.}
\label{table:mwmodel}
\begin{center}
\begin{tabular}{c|ccccc}
\hline
Component & Fraction & Scale Height & Velocity Model & Velocity Mean & Velocity Dispersion\\
 & & pc & & $\kms$ & $\kms$ \\ 
\hline
young disk & 0.250 & 76 & Dehnen & $0,-10,0$ & $15,15,10$ \\
\hline
thin disk & 0.650 & 320 & Dehnen & $0,-20,0$ &  $40, 35, 20$ \\
\hline
thick disk & 0.100 & 800 & Dehnen & $0,-30,0$ & $55, 55, 40$ \\
\hline
underlying halo & 0.00518 & 6000 & Gaussian & $0,-220,0$ & $140,100,100$\\
\hline
{\it Gaia}-Enceladus-Sausage & 0.00155 & 6000 & double-Gaussian &  $\pm150,-240,-10$ & $90,40,60$ \\
 & & & mixture & & \\
\hline
\end{tabular}
\end{center}
\end{table*}

\subsection{Determination of tangential velocity cut and correction factor for measuring the stellar halo LF}

\begin{figure*}
\includegraphics[width=\textwidth]{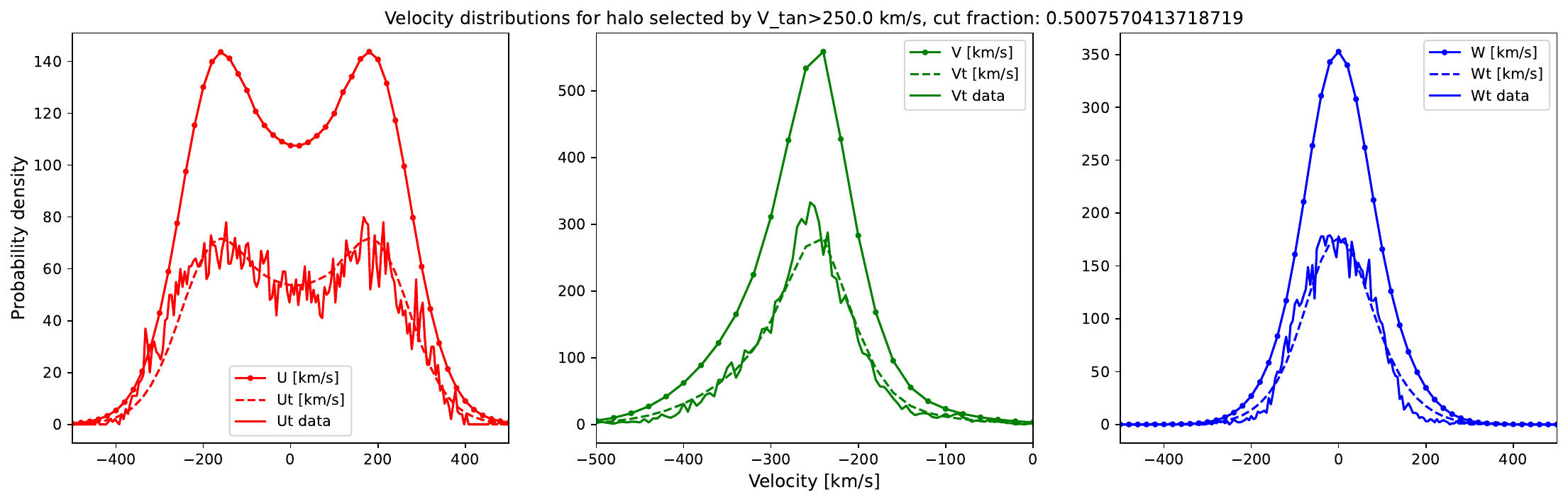}
\caption{Cartesian velocity $(U,V,W)$ (left, middle, right panels, respectively) histograms comparing our model (connected dot curve), our model after applying a transverse velocity cut of $V_\mathrm{t}>250\,\kms$ (dashed curve), and our transverse velocity selected halo stars with {\it Gaia} radial velocity (solid line).
}
\label{fig:uvw}
\end{figure*}

From Fig. \ref{fig:mwmodel}, uppermost panel, we see that $V_\mathrm{t}>250\,\kms$ is an appropriate cut to retain a pure halo sample, excluding thin and thick disk components.

Using such a selection excludes half of the total number of halo stars as these have similar transverse velocities as disk stars. 

To estimate the stars lost due to our $V_\mathrm{t}$ cut of 250\,$\kms$ we use the halo components of our Milky Way model to investigate the halo distribution function 
in 3D velocity space within the solar neighborhood using the two components, one representing the underlying halo and the second the {\it Gaia}-Enceladus-Sausage satellite \citep[][]{Belokurov2018,Deason2018.862,Haywood2018,Helmi2018,Koppelman2018}.

We model the 1 kpc spherical volume of halo stars with the cut in Galactic latitude $b>36^\circ$. We then apply the transverse velocity cut $V_\mathrm{t}>250\,\kms$ and compare the resulting velocity distribution to the cartesian $(U,V,W)$ velocities of the halo stars within our sample that have measured {\it Gaia} radial velocities in addition to the proper motion and parallax measurements.

In Fig. \ref{fig:uvw} we show our model and the data. The model and data curves both with transverse velocity cuts (dashed and solid histograms) are comparable giving us confidence that our model is adequate for calculating the number of missing halo stars due to the transverse velocity cut. We find a correction factor of 2.0 and we apply this in our LF calculations.

\end{document}